\documentclass[a4paper, 10pt, conference]{ieeeconf}      

\IEEEoverridecommandlockouts                              
\usepackage{graphicx} 

\title{\LARGE \bf
HORST: Holographic Orbital Return Storage Technology\\
\large \bf www.horst.space
}

\author{Matthias Raudonis$^*$, Natalia Triantafyllou, Anna Zhuravlova\\
\small matthias@raudonis.space, nataliatriantafyllou@gmail.com, annazhuravleva7@gmail.com%
}

\begin{document}

\maketitle
\thispagestyle{empty}
\pagestyle{empty}

\begin{abstract}		
Nowadays, space science is facing increasing problems with the amount of data collected from sensors in space and its transmission back to Earth.
In this paper we introduce the novel Holographic Orbital Return Storage Technology (HORST) and its potential application in space industry.
The proposed solution is a payload module which stores hundreds of terabytes of data on a robust 5D holographic disk. After the end of mission (EOM), the module is detached from the satellite and lowered into the Earth's atmosphere, protected by a heatshield surface and a parachute. 
The recovery of the module allows the readout of big sensor data on Earth.
Besides fulfilling the big demand for applications of this technology nowadays, this paper discusses several major use-cases for near-future concepts and missions.
HORST will enable many possibilities for new science missions and business in space. Since there is no comparable alternative technology, the lack of competition and the increasing demand will allow HORST to become a key technology for space.

\end{abstract}

\section{INTRODUCTION}
\label{sec_intr}
In 1965, Gordon Moore accurately predicted that the number of components
on an IC would double every year for the next 10 years. 53 years later (2018), this rule, which is later known as Moore's Law, is still applicable. \cite{c2} 
With the increase in processing power and sensor accuracy, the amount of data measured, processed and stored is continuously growing -- on earth and in space.
For data transmission, most satellites rely on the electromagnetic microwave (L, S, C, X, Ku, Ka) bands, where insufficient transmission bandwidth and congestion is already a problem as of today. \cite{c3}
To tackle this bottleneck for the present and near-future, this paper introduces the concept of a payload module to physically return data back to earth, instead of RF transmission.

\section{HOLOGRAPHIC DATA STORAGE}
\label{sec_holst}

   \begin{figure}[htpb]
      \centering
      \includegraphics [width=.5\textwidth] {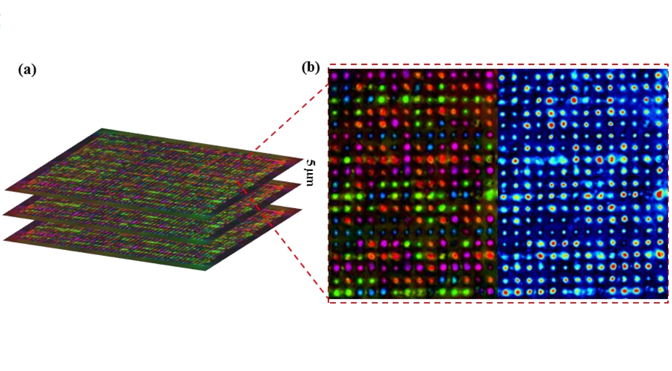}
      \caption{Holographic 5D stored data matrix \cite{holo_bib}}
      \label{holo_fig_lab}
   \end{figure} 

Nowadays, two-dimensional data storage technologies (CD, DVD, Blu-ray Disk) are common which use only one layer to store data.
The idea of the optical memory based on a femtosecond laser, writing in the bulk of transparent material, was proposed in 1996 and implemented in 2013 in Optoelectronics Research Centre, University of Southampton, where a 5D disk was created. In this case two more dimensions for data storing are added apart of the three spatial coordinates: The slow axis orientation (4th dimension) and the strength of retardance (5th dimension). (see figure \ref{holo_fig_lab})
The more dimensions, the more data can be stored in the same volume.
This technology allows to store 360 TB/disk! 
Nanostructured silica glass disks are used, because this material is resistant to rapid heat changes, mechanical shock and aggressive radiation. To change the structure of the glass to store data, a femtosecond laser is required. This technology allows to store data without degradation for hundreds of years and an infinite amount of read-cycles. [8]

\section{PAYLOAD MODULE}

The payload module itself is a round capsule with an outer ablative heat shield layer to protect the module from atmospheric reentry thermal pulse and mechanical shock. 
Inside the capsule, the module has stored a commercial off-the-shelf (COTS) parachute to enable a soft landing. The separation mechanism of the upper part of the heat shield hull to expose the parachute is critical and has to be developed and tested.
Standard control electronics and a serial data interface like SpaceWire (LVDS) or CAN are also necessary, to connect the payload module with the satellite and to convey data further to the holographic laser writer. 
Also, a transponder for recovery tracking is necessary for the recovery. The density of the module shall be less than $1020 \frac {kg}{m^3}$ because it has to float on sea water.

The main system, enclosed in the bottom of the module, is the holographic laser writer, which includes a laser with one or several wavelengths and a set of lenses and mirrors to focus the laser beam. It allows to write data on one or multiple nanostructured silica glass disks. The disks and the laser have to be bedded in a shock-absorbing supporting mount to increase the mechanical shock resistance.

\section{DATA RETURN TO EARTH}
\label{sec_dat_ret}

\subsection{Target Return Orbit}
Before the separation of the HORST payload module, the satellite must be in an very low earth orbit to initiate aerobraking to bring the module down to earth, since the module itself doesn't contain any means of propulsion to lower the orbit. The orbit may be elliptical in case of soft aerobraking. However, the module is not designed for a steep, direct descent due to heat shield requirements.
After detaching, the satellite could rise to a stable orbit again or be decommissioned into the atmosphere, depending on the mission goal.
\subsection{Recovery}
After the payload enters the atmosphere, a parachute will deploy to slow down its speed for landing. The landing itself will take place in a defined dropping zone, a sea area close to the shore. Its range is directed by many variables and include among others the orbit and the shape of the payload, as well as dynamic parameters like the weather conditions and jet streams. After landing, the exact location will be known due to a tracking device, which is included in the module. A contractor ship will follow the signal and retrieve the payload.   
\section{USE CASES}
\label{sec_useC}
This new, versatile technology has many possible use-cases. In this section, we identified the main present to near-future use cases for the HORST payload module.


   \begin{figure}[htpb]
      \centering
      \includegraphics [width=.5\textwidth] {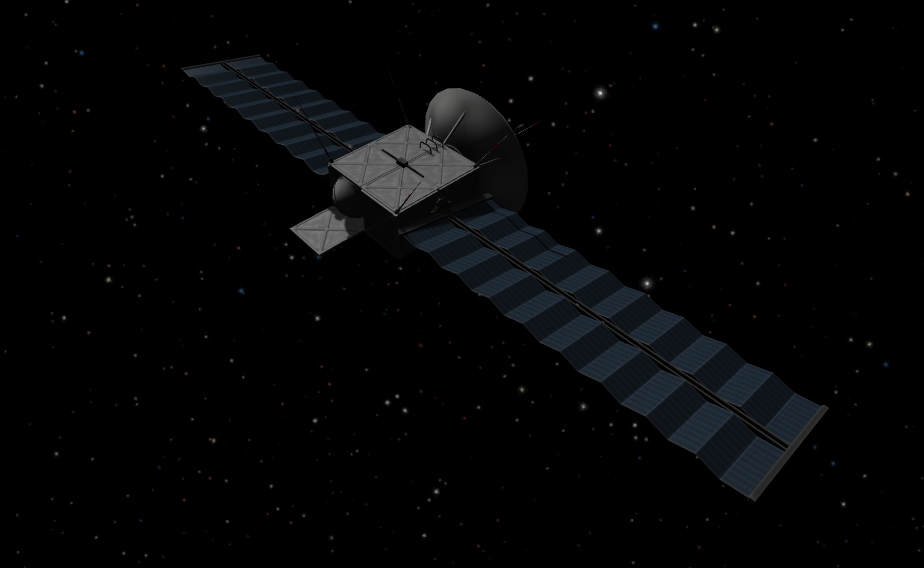}
      \caption{Satellite visualization with HORST module in payload bay \cite{ksp}}
      \label{figuresat}
   \end{figure}
 
\subsection{Safe Data Storage}
An alternative use case is the safe storage of data in space. The data would be slowly transmitted by existing microwave RF systems. The properties of the holographic storage, mentioned in \ref{sec_holst} make it an optimal safe long term data storage.
In case of a major catastrophe on earth, the data will still be safe in orbit. 
This increases data redundancy for safe cloud storage architectures.\\
In February 2018, SpaceX launched a Falcon Heavy to a solar orbit, carrying a disk with this technology to safely store and preserve libraries of human knowledge. \cite{spacexholo}

\subsection{Data Transmission}
For mid-future human Lunar or Martian bases, essential and up-to-date data can be slowly transmitted by existing microwave RF systems.
However, big amounts of research, entertainment or informational data can be transferred from and to earth by the HORST module. 
For example, a resupply mission to a Lunar or Martian base could contain a HORST module with movies and music for entertainment, recent news and an incremental update of an encyclopedia like Wikipedia. 

\subsection{Space Very-long-baseline Interferometry}
\label{sec_vlbi}
   
The main present to near future use case is the use of Very-long-baseline interferometry in a space satellite constellation (S-VLBI). \\
\subsubsection{Satellite Constellation with HORST Payload}
\label{sec_const}

   \begin{figure}[htpb]
      \centering
      \includegraphics [width=.5\textwidth] {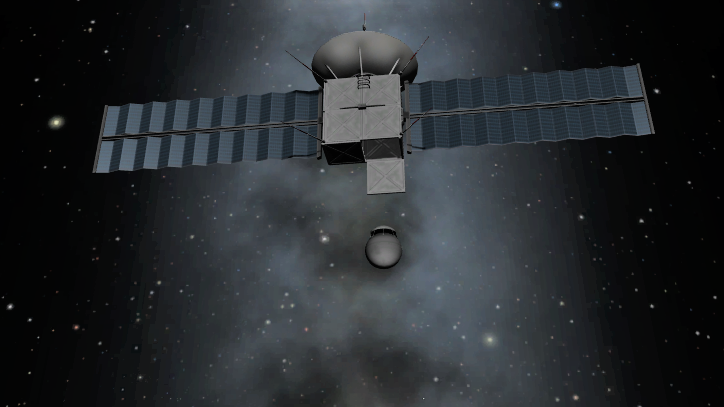}
      \caption{S-VLBI usecase satellite visualization with separated HORST module \cite{ksp}}
      \label{figuresat_sep}
   \end{figure} 
   
With this technology, multiple satellites can build up a virtual radio telescope with a ultra high angular resolution. A longer distance between the individual radio telescopes translates into a higher angular resolution. \cite{EHT} Therefore, the satellites shall be injected into a high earth orbit, where the distances between the satellites and earth is in the magnitude of tens of thousands of kilometers, which is one of the main advantages of S-VLBI.\\
This method of VLBI works by accurately sampling digital telescope data with the reference of an atomic clock and saving it onto the HORST payload module. At the EOM, the satellite orbit is lowered for reentry and the HORST module is separated. After the recovery of the HORST modules with all the constellation data, a high resolution image can be reconstructed on earth by a computing cluster.\\
Following payload modules are proposed:
\begin{itemize}
\item Radio telescope dish and receiver
\item Accurate atomic reference clock (H maser), can be used from ESA's GALILEO Mission
\item RF Transceiver for calibration and satellite control
\item HORST Module for big storage data return\\
\end{itemize} 
\subsubsection{Alternative Data Gateway Topology}
Instead of using one HORST module per radio telescope satellite, the proposed topology in section \ref{sec_datasink} can facilitate the data transfer. The major benefits are longer mission times, faster data availability and simpler satellite design, which leads to a cost advantage.
\subsubsection{Current development and projections}
In 2017, the Event Horizon Telescope (EHT) used VLBI with multiple radio telescopes around the world to create a virtual telescope of the size of the earth with the purpose of rendering an image of a black hole's event horizon. Since the EHT also uses decentral data storage and physical transfer, the data readout and processing is not complete yet. However, the first actual image of a black hole is expected by the end of the year 2018 which could reveal what no telescope was able to resolve yet. \cite{EHTstatus} The proposed Space VLBI constellation in \ref{sec_const} with a virtual telescope of the size of hundreds of earths will be powerful and accurate enough to reveal many secrets of our universe.


Through the use of existing technology, like the GALILEO atomic reference clocks, and the production of a satellite constellation based on the existing satellite platforms, the proposed VLBI telescope can be implemented with a limited budged while providing previously unreached accuracy and resolution.


\subsection{Optical Space Transmission Star Topology Data Sink Gateway}
\label{sec_datasink}
In 2001, ESA achieved the first satellite to satellite laser transmission (Artemis satellite). Because of the shorter wavelengths of light and the possibility to use multi-chromatic light, very high data rates can be achieved. \cite{esa_trans}\\
To serve the data storage and transmission needs of multiple satellites, one or more satellite nodes with HORST modules and an optical laser transmission system can be positioned in the Low Earth Orbit (LEO). 
Multiple satellites can upload sensor data to these satellite nodes via high speed optical links. After a holographic disk on a node is full, the HORST Module can be detached to go back to earth, as described in section \ref{sec_dat_ret}. \\
This system has the major advantage that the data acquisition systems is physically separated from the data storage and transmission system (HORST). 
This enables long satellite observation or exploration missions while the HORST node satellite detaches the modules more frequently for shorter data transmission times.\cite{basc}
Examples for possible missions are discussed in sections \ref{sec_vlbi} and \ref{sec_explo}.
\subsection{Exploration and observation missions}
\label{sec_explo}
The data return principle used in section \ref{sec_vlbi} can also be implemented in other exploration and observation satellite payload missions. With the massive increase of available data storage capabilities, new sensors can start fully using their accuracy and resolution potential without the limitation of the downlink bottleneck, mentioned in \ref{sec_intr}.
This will yield many new scientific discoveries, mainly in astronomy.

\section{IMPLEMENTATION STRATEGY}

They key activities of HORST is the provision of a huge capacity of storage space, access to data in high resolution while ensuring their safety and stability. Regarding the analysis in the section \ref{sec_useC}, HORST's business case is mainly B2B (business to business). The main customers are from the scientific sector. 

Because the module is brought back to earth, many components will still be functional upon recovery. The implementation goal is complete re-usability, apart from the full returned holographic disk.

\section{CONCLUSION}
The holographic orbital return storage technology is the solution for huge data storing and delivery back to Earth, as this storage is dense (much bigger capacity comparing to technologies used now), resistant to time (stable over hundreds of years), radiation, temperature change and mechanical shock, what makes it a promising technology for future scientific and private sector use. By re-using the module and the use of COTS parts, the module is very cost effective.
It will be useful for industry both in space and on Earth, as data is a key resource in our digital world.
By working on the bandwidth bottleneck problem, described in \ref{sec_intr}, we created the concept of a device with many applications. There is no comparable alternative technology. The five use-cases discussed in section \ref{sec_useC} enable new, daring science and business in space and on earth, but are only a fraction of what is and what will be possible.


\flushleft

Also published on www.horst.space\\ \today, Version 1.1

\end{document}